\documentclass[12pt]{iopart}

\usepackage{graphicx}
\usepackage[dvipsnames]{xcolor}

\makeatletter
\newcommand{\thickhline}{%
    \noalign {\ifnum 0=`}\fi \hrule height 1pt
    \futurelet \reserved@a \@xhline
}

\begin{document}
\title{An autoencoder for compressing angle-resolved photoemission spectroscopy data}

\author{Steinn {\'Y}mir {\'A}g{\'u}stsson}
\address{Department of Physics and Astronomy, Aarhus University, 8000 Aarhus C, Denmark}

\author{Mohammad Ahsanul Haque}
\address{Department of Computer Science, Aarhus University, 8000 Aarhus C, Denmark}

\author{Thi Tam Truong}
\address{Department of Computer Science, Aarhus University, 8000 Aarhus C, Denmark}

\author{Marco Bianchi}
\address{Department of Physics and Astronomy, Aarhus University, 8000 Aarhus C, Denmark}

\author{Nikita Klyuchnikov}
\address{Independent researcher, Dubai, United Arab Emirates}

\author{Davide Mottin}
\address{Department of Computer Science, Aarhus University, 8000 Aarhus C, Denmark}

\author{Panagiotis Karras}
\address{Department of Computer Science, Aarhus University, 8000 Aarhus C, Denmark}

\author{Philip Hofmann}
\address{Department of Physics and Astronomy, Aarhus University, 8000 Aarhus C, Denmark}

\date{\today}
\begin{abstract}
Angle-resolved photoemission spectroscopy (ARPES) is a powerful experimental technique to determine the electronic structure of solids. Advances in light sources for ARPES experiments are currently leading to a vast increase of data acquisition rates and data quantity. On the other hand, access time to the most advanced ARPES instruments remains strictly limited, calling for fast, effective, and on-the-fly data analysis tools to exploit this time. In response to this need, we introduce ARPESNet, a versatile autoencoder network that efficiently summmarizes and compresses ARPES datasets. We train ARPESNet on a large and varied dataset of 2-dimensional ARPES data extracted by cutting standard 3-dimensional ARPES datasets along random directions in $\mathbf{k}$. To test the data representation capacity of ARPESNet, we compare $k$-means clustering quality between data compressed by ARPESNet, data compressed by discrete cosine transform, and raw data, at different noise levels. ARPESNet data excels in clustering quality despite its high compression ratio.
\end{abstract}
\maketitle

\section{Introduction}

Angle-resolved photoemission spectroscopy (ARPES) is an experimental technique to determine the electronic structure of crystalline solids based on the photoelectric effect: X-ray photons of energy $h\nu$ hit the surface of a material, causing the emission of photoelectrons. The resulting photoemission current (or photoemission intensity) $I$ can be detected by a photoelectron spectrometer, resolving the electrons' kinetic energy and two emission angles~$\Theta$ and~$\Phi$ (and, in some experiments, the electron spin). From these quantities, it is possible to determine the binding energy of the electrons and the two-dimensional (2D) crystal momentum $\mathbf{k}=(k_x, k_y)$ parallel to the surface and, with some additional assumptions, also the component of the crystal momentum perpendicular to the surface~$k_z$. Importantly, the photoemission intensity as a function of energy and crystal momentum~$\mathbf{k}$ is closely related to the sample's spectral function, a quantity that is of key importance for many properties such as the solids' conductivity, possible superconductivity, band topology and others~\cite{Sobota:2021wy}. ARPES measurements can unravel the intricate electronic properties of quantum materials and the many-body interactions in such materials~\cite{Damascelli:2003aa, Hofmann:2009ab}. 

The typical result of a modern ARPES measurement is an image, or \emph{spectrum}, of the photoemission intensity as a function of kinetic energy and one emission angle (or $\mathbf{k}$ in one direction), while the other angle needs to be explored by sample rotation. Such a raw data image is shown as input to the autoencoder network in Fig.~\ref{fig:1}. For a more detailed description of ARPES, see, e.g. Refs.~\cite{Sobota:2021wy, Damascelli:2003aa, Eberhardt:1980aa,Hofmann:2021tj}

Recent progress in X-ray light sources has lead to a surge in ARPES capabilities. In particular, the ability to produce ultrashort light pulses by high-harmonic lasers and free electron lasers has opened the possibility to study non-equilibrium phenomena on an ultrafast time scale, leading to the development of time-resolved ARPES~\cite{Boschini:2024aa}. Also, modern synchrotron radiation sources allow the X-ray light spot to be tightly focused to below 1$\mu$m such that photoelectrons are only emitted from a very small region of the sample. Scanning this light spot across the sample surface turns ARPES into a microscopy technique where a full dataset can be acquired for every desired position on the sample. Such position-resolved experiments are called microARPES or nanoARPES, depending on the spatial resolution \cite{Rotenberg:2014aa, Ulstrup:2019aa,Ulstrup:2020aa}. It has been possible to expand such experiments to investigating electronic devices and non-equilibrium situations such as electrostatic doping or the presence of transport currents~\cite{Nguyen:2019aa, Curcio:2020aa, Curcio:2023aa, Hofmann:2021tj}.

These developments have greatly increased the parameter space of photoemission experiments. The basic building block remains a collection of 2D images as in Fig.~\ref{fig:1}, but images are now taken as a function of time, position on the sample, temperature, device gate voltage, transport current, and so on. Exploring such a high-dimensional parameter space by collecting data for all parameter combinations is prohibitively time-consuming, calling for efficient techniques to determine ``interesting'' regions of parameter space, such as Gaussian process regression~\cite{Noack:2021aa, Agustsson:2024aa}. Nevertheless, it often remains necessary to collect many spectra and therefore compromise in terms of counting statistics, resulting in ARPES images of a low signal-to-noise ratio ($S/N$). Lastly, on-the-fly data analysis is called for to detect patterns in the data collected up to a point in an experiment, to make informed decisions on the progress of that experiment.

An extremely useful method for pattern detection in complex ARPES data is clustering~\cite{Jain:1999aa}, as for example realised by the $k$-means algorithm~\cite{Bock:2007wz, Melton:2020aa} that employs a Euclidean distance measure among photoemission intensity spectra. The approach is particularly well-suited for the analysis of ARPES data to determine how many areas of fundamentally different properties exist on a surface and where they are~\cite{Iwasawa:2022uw, Mortensen:2023aa}. However, calculations of Euclidean distance among ARPES spectra are demanding, as each pixel of the spectrum enters the calculation as a different dimension or feature, and hence they benefit from efficient formulations of $k$-means~\cite{Mortensen:2023aa}.

An \emph{autoencoder} as the one in Fig. \ref{fig:1} encodes the input data---a photoemission spectrum---to a lower-dimensional latent-space representation and then decodes this again to the same size as the input image.
The network is trained by exhorting the output to (i)~resemble the input as closely as possible and (ii)~generalise to unseen or noisy input examples.
 A successful outcome implies that the salient features of the data are also contained in the compressed representation in latent space.  This representation can then be used to extract important properties of the data or, as in this paper, serve as an input to clustering by $k$-means.
 Using the compressed version of the data therefore supports fast on-the-fly data analysis, while the time-consuming compression can be performed concurrently with the collection of additional ARPES images.

In this article, we craft an autoencoder network for the efficient compression of ARPES data. After training this network, we explore the quality of~$k$-means clustering on the compressed latent-space representations. To this end, we construct a test dataset of noisy ARPES images derived from a known ground truth and perform $k$-means clustering using the latent-space representation generated by the autoencoder. Operations on the compressed data outperform those on the raw data not only in terms of speed, but also in terms of clustering quality for a wide range of $S/N$. We confirm this superior performance by applying autoencoder compression and~$k$-means clustering to published photoemission data from a graphene device~\cite{Curcio:2020aa}.

\section{Autoencoder Network}

In the context of ARPES and other general spectral techniques, neural networks can serve various purposes such as denoising, resolution enhancement, and feature recognition. The former two aspects have received some attention and have been implemented primarily using convolutional neural networks \cite{Kim:2021vx,Peng:2020aa,Huang:2024ab}. Here we focus on data summarisation and compression, hence we employ an autoencoder network.

\subsection{ARPESNet Architecture}

We introduce ARPESNet, a deep convolutional autoencoder. ARPESNet consists of an encoder responsible for compressing the input and a decoder tasked with reconstructing the input based on the compressed representation. Contrary to typical image compression methods such as JPEG~\cite{Panchanathan:1996aa}, autoencoders  reduce the \emph{dimensionality} of the input image rather than its bit-size representation. We do not use a quantiser after the encoder as is otherwise common in image compression techniques.

\begin{figure}
\includegraphics[width=1.0\columnwidth]{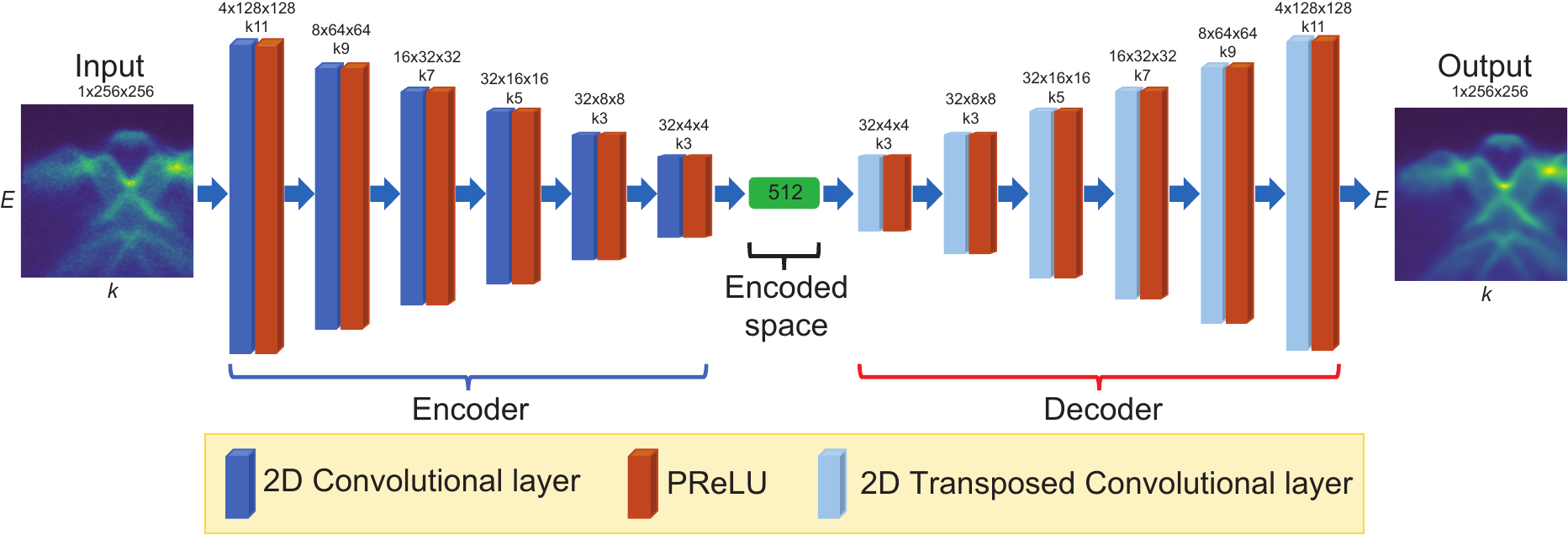}
\caption{Structure of the ARPESNet autoencoder. The input and output are ARPES images of photoemission intensity as a function of crystal momentum~$k$ and energy~$E$. The labels on top of the compression and decompression blocks are structured as~``$c$$ \times $$a$$ \times$$b$k$N_k$", where~$c$, $a$ and~$b$ indicate the number channels and lateral sizes respectively, and~$N_k$ is the kernel size. In the encoder (decoder) block, blue (light blue) bars indicate (transposed) convolutional layers, while red bars indicate parametric rectified linear unit (PReLU) activation functions~\cite{Kaiming:2018aa}.}\label{fig:1}
\end{figure}

Figure~\ref{fig:1} shows the structure of ARPESNet. Encoder and decoder are built as mirrors of each other, with~6 compression blocks made of a convolutional layer and an activation function in the encoder. The decoder has~6 decompression blocks, substituting the~2D convolutional layer with a~2D transposed convolutional layer. We chose parametric rectified linear unit (PReLU) activation functions~\cite{Kaiming:2018aa}, which stabilise training by adding a learnable parameter to the standard ReLU activation functions. The first compression block in the encoder increases the channel number from~1 to~4, and each subsequent block increases it further by a factor~2, up to a maximum of~32 channels. The kernel size is set to~11 for the initial block and decreases by~2 in each subsequent block to a minimum of~3. The decompression blocks of the decoder mirror the encoder, starting from~32 channels with kernel size~3 and ending with~1 channel and kernel size~11. To ensure conservation of the image shape upon reconstruction, all convolutional layers apply zero-padding of~$p = (N_k - 1)/2$ pixels, where~$N_k$ is the layer's kernel size. Likewise, the transposed convolutional layers have an output padding of 1, matching the stride of~2 in each layer. This architecture aims to capture both large- and small-scale features in complex ARPES data. 
The resulting network has~$81,389$ parameters and compresses a~$256 \times 256$ image to an encoded space of size~$512$. This, in turn, is reconstructed to a~$256 \times 256$ image by the decoder. The compression ratio (CR) is thus~${256^2}/{512} = 128$.
The hyperparameters were optimised by minimising the training loss after 100 epochs using the Optuna python package~\cite{Akiba:2019aa}.

\subsection{Training Data} 

\begin{figure}[!t]
\includegraphics[width=1.0\columnwidth]{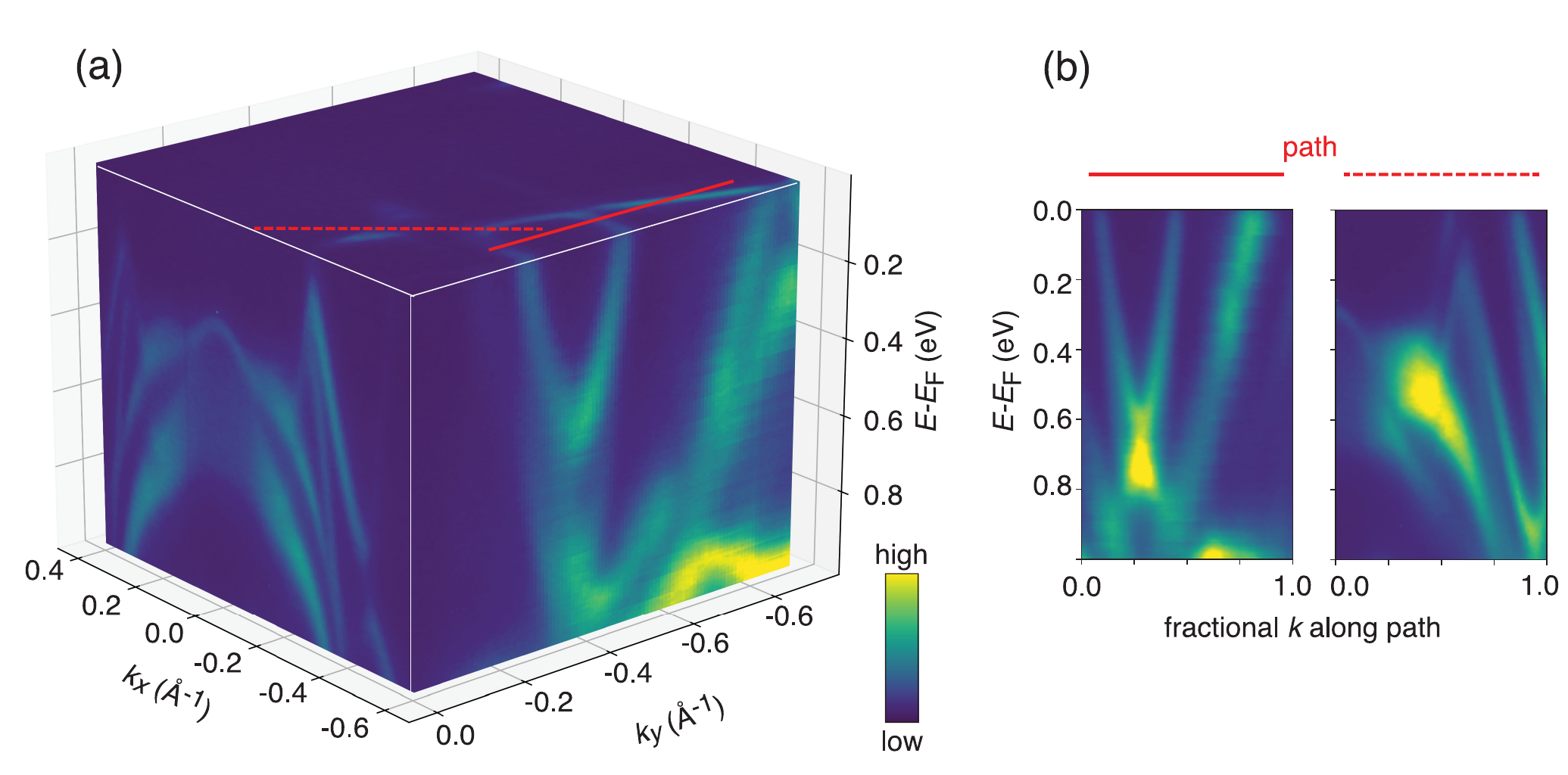}
\vspace{-2mm}
\caption{Extraction of training images from a 3D ARPES dataset. (a)~A 3D dataset for NdTe$_3$~\cite{Chikina:2023aa}; solid and dashed red lines mark random cuts in~$k_x, k_y$ defining 2D ARPES spectra. (b)~Photoemission intensity images along these two paths.}
\label{fig:2}
\end{figure}

The quality and variety of training data are essential for network performance. Here, we introduce a procedure to extract numerous training images from a standard ARPES experiment. The output of a~2D ARPES detector is an image of photoemission intensity as a function of electron kinetic energy~$E$ and one emission angle~$\Theta$, i.e.,~$I(E,\Theta)$. Typically, one measures a series of such images by varying the other emission angle~$\Phi$ in small steps to assemble a three dimensional (3D) dataset~$I(E, \Theta, \Phi)$. Fig.~\ref{fig:2}(a) shows such a dataset, with the slight modification of converting the angles~$\Phi$ and~$\Theta$ into crystal momentum components~$k_x$ and~$k_y$, respectively~\cite{Eberhardt:1980aa}, resulting in~$I(E, k_x, k_y)$. The training dataset could now be selected from the primary~$I(E, \Theta)$ or, equivalently,~$I(E,k_x)$ images; however, that would be a poor choice due to the high degree of symmetry caused by the sample alignment prior to measurement, which induces a bias towards symmetric images, the high similarity between neighbouring images in a~$\Phi$-scan, and the scarcity of images in total, typically on the order of~100.

To increase the variety and size of the training dataset, we employ the following data extraction method: Starting from a dataset such as in Fig.~\ref{fig:2}(a), we define a random path of fixed length in~$k_x, k_y$ space. These two dimensions represent the same physical quantity (crystal momentum) at the same scale and unit (m$^{-1}/$pixel), so that any line in the~($k_x$, $k_y$) plane has the same units. The solid and dashed red lines in Fig.~\ref{fig:2}(a) show two such paths. Fig.~\ref{fig:2}(b) depicts the extracted photoemission spectra for these two paths. In the case of metallic samples, such as the one in Fig.~\ref{fig:2}, the measured energy range typically extends above the Fermi energy~$E_\mathrm{F}$. In Fig.~\ref{fig:2} we omit data above~$E_\mathrm{F}$ for illustrative purposes but include them in training data. By selecting random paths through the data, we extract hundreds of rather different ARPES images from a single~3D dataset, while carefully limiting the energy and~$k$-range so that the number of  bands / structures in the data is low enough to be represented in an image of the set resolution (e.g.,~$256 \times 256$). We revisit the resolution issue with a more detailed discussion below.

We apply this procedure to 46 ARPES data sets from 19 different materials. From these, 28 datasets covering all materials are selected to generate training data. The remaining 18 datasets, covering 12 materials are used to generate a test dataset.
We apply this procedure to~46 ARPES datasets from~19 different materials. Out of these, we select~28 datasets covering all materials as training data and the remaining~18 datasets, covering~12 materials, as test data. While some materials are present in both the training and test data, we take care to ensure that the images show other spectral features, which render them sufficiently different to be considered independent (as images). We generate~500 images from each dataset, yielding~14,000 and~9,000 training and test images, respectively. Repositories of suitable ARPES data, as well as the complete dataset we have used for training ARPESNet \cite{Sanders:2016aa, Dendzik:2017ab, Holt:2020aa, Holt:2021vs, Chikina:2023aa, Dreher:2021wy, Miwa:2015aa, Ulstrup:2018aa, Larciprete:2012aa, Bianchi:2023aa, Bianchi:2012ab, Bianchi:2010ab, Bianchi:2011aa, Dendzik:2016aa}, are published on Zenodo (DOI: 10.5281/zenodo.12648783 and 10.5281/zenodo.12665275). We specifically choose some photoemission images from the test data for visual inspection, to examine ``worst cases'' in which a high-quality reconstruction was especially difficult. The spectra in Figure~\ref{fig:3} are a selection of such worst cases.

To monitor stability during training, we split the~14,000 image dataset into training and validation parts at an~80:20 ratio. We further augment the data by cropping random regions from the images. The size of the cropped regions is randomly chosen to be between 80\% and 100\% of the original size, with the aspect ratio of the crop randomly adjusted between 0.8 and 1.2. These images are flipped horizontally with 50\% probability. Rotations are not allowed to ensure that the images retain their physical meaning where the horizontal and vertical axes correspond to~$k$ and energy,  respectively. Lastly, intensity is normalsed to the interval~$\left[0,100\right]$.

\subsection{Training and Testing}

We train ARPESNet aiming to achieve the best possible agreement between the output images of the network and suitably selected reference images. Input and reference images are generated for each instance in the training data. While the images in the training dataset are of very high quality, with a low~$S/N$, it is desirable to also train the network using noisy images resembling a realistic experimental situation. We create such noisy images by resampling the high-quality training images using Poissonian statistics to simulate a  number of collected electrons~$n_I$ or, equivalently, a certain acquisition time. We explore values of~$n_I$ between~$10^4$ (low $S/N$ ) and~$10^8$ (high $S/N$). Figs.~\ref{fig:5}(f)-(i) show examples of images resampled in this way. 

We tried different combinations of input and reference images to evaluate loss while training ARPESNet. First, we simply used the high-quality image for each instance in the training data as input to ARPESNet and compared the output to the same image. We refer to this training configuration as \emph{no-noise}. We also produced a noisy version of the reference image to serve as input data, as described previously, and compared the output image either to the generating high-quality image or to the same noisy image. We consider the former case to train ARPESNet as a denoising network, hence dub it \emph{denoiser}; we found the latter, \emph{noise-to-noise} training to yield excellent denoising for pixelated images at low count numbers, as long as it is trained with many different noisy images.

\begin{table}[htbp]
  \centering
  \footnotesize
  \begin{tabular}{lll|rr|rr}
    \multicolumn{3}{r}{} & \multicolumn{2}{c}{original} & \multicolumn{2}{c}{$n_I=10^5$} \\
    model    & loss           & epochs & MSE  & PSNR  & MSE  & PSNR   \\
   \thickhline
    ARPESNet & no-noise         & 1,000  & 0.48 & 40.12 & 1.53 & 21.78  \\
    ARPESNet & noise-to-noise & 1,000  & 0.40 & 41.12 & 1.52 & 21.80  \\
    ARPESNet & denoiser       & 1,000  & 5.53 & 28.24 & 1.79 & 21.11  \\
    ARPESNet & noise-to-noise & 4,000  & \textbf{0.38} & \textbf{41.34} & \textbf{1.52} & \textbf{21.80}  \\
    \hline
    DCT22    &                &        & 1.01 & 38.63 & 1.54 & 21.77  \\
    \end{tabular}
\caption{Mean mean square error (MSE) and peak signal-to-noise ratio (PSNR) values evaluated on 9,000 test spectra for different combinations of training and test strategies for ARPESNet. The first four rows represent different training loss configurations, each trained for 1,000 epochs, as well as the noise-to-noise training configuration, trained for 4,000 epochs (results in bold). The columns to the right show the two metrics evaluated on the original spectra and on noisy spectra with $n_I=10^5$. The last row gives the corresponding results for DCT with a similar compression rate.}
\label{tab:network_performance}
\end{table}

We train ARPESNet using the Adam optimiser~\cite{Kingma:2014aa} with a fixed learning rate of~$lr = 0.001$ and no weight decay, using mean square error (MSE) as a loss function. Initially, we train all input/reference combinations (no-noise, noise-to-noise, denoiser) for~1,000 epochs. Tests on the~9,000 test images produce the results shown in Tab.~\ref{tab:network_performance} using MSE and peak signal-to-noise ratio (PSNR). These performance indicators, as well as a preliminary test of the $k$-means clustering performance, show a clear preference for the noise-to-noise model, which we subsequently train for another~20,000 epochs. We obtain the lowest MSE at~4,000 epochs, steadily worsening thereafter due to overfitting. We exclusively use this model trained for 4,000 epochs in the results section. We also compare results to those of a discrete cosine transform (DCT)~\cite{Cintra:2011aa, Dimililer:2022aa, Mortensen:2023aa} representation truncated to the first~$22\times22$ pixels, which we refer to as DCT22 in Tab.~\ref{tab:network_performance} and in the following; DCT22 yields a compression ratio of~$CR = 135.4$, close to the one of ARPESNet ($CR=128$).

\begin{figure}
\includegraphics[width=1.0\columnwidth]{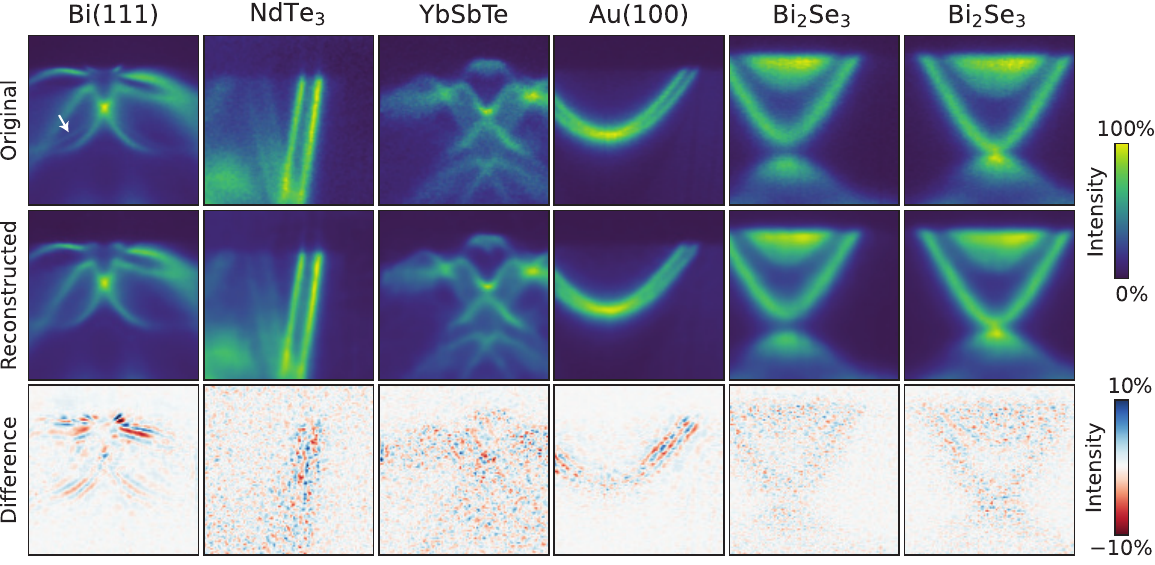} 
\caption{Testing the performance ARPESNet after 4,000 epochs of training. Top row: Test images; middle row: reconstruction and bottom row: normalised residual. All images show the photoemission intensity as a function of energy (vertical) and crystal momentum (horizontal). The arrow in the original image of the first column marks a weak but sharp state that is not well-captured in the reconstruction.}\label{fig:3}
\end{figure}

The reconstruction quality of ARPESnet is excellent; Fig.~\ref{fig:3} gives a visual impression thereof. Reconstructions are usually even better than those in the figure, as the images in Fig.~\ref{fig:3} are chosen from the ``worst case'' scenarios. The difficulties arising with these particular images are revealed by a careful inspection of the normalised residual in the third row. For instance, in the first column, the original shows two tiny electronic states near the top-middle of the images (the electron pocket of the Rashba-split surface state in Bi(111)~\cite{Koroteev:2004aa}), as well as sharp lines to the left and right below the pronounced and bright crossing point in the middle of the image (marked by an arrow). These are not well-reproduced in the reconstruction and the corresponding regions of~$(E,k)$ space show a relatively high residual.

\begin{figure}[htbp]
\centering
\includegraphics[width=0.8\columnwidth]{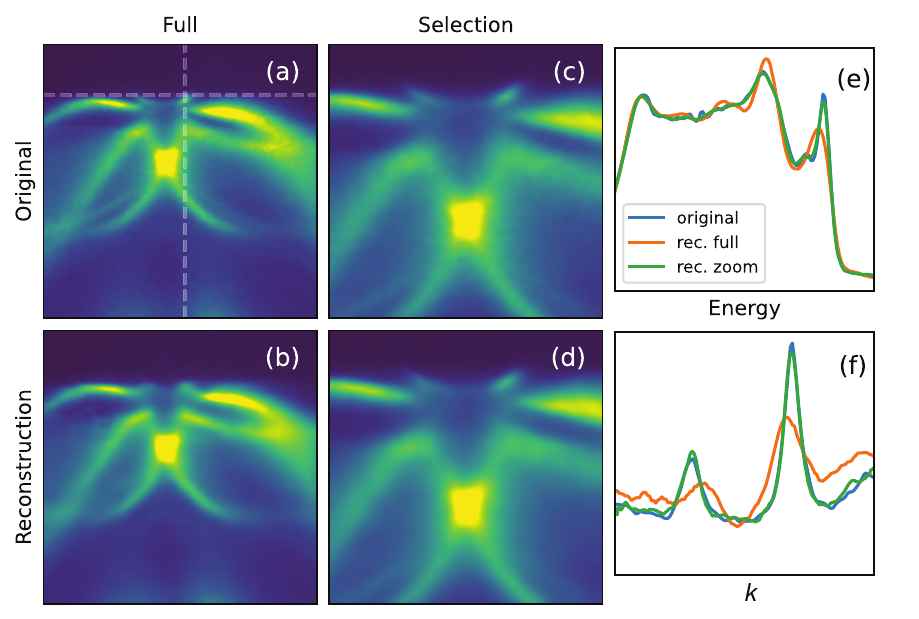}
\caption{Shortcomings of reconstructing sharp features in the test data of Fig.~\ref{fig:3} are solved by choosing a different scaling. (a) Data for the Bi(111) from the top row, first column of Fig. \ref{fig:3}. The dashed lines indicate the paths for the photoemission intensity cuts in panels e and f. (b) Reconstruction of the data in panel a. (c) Zoomed-in version of the data. (d) Corresponding reconstruction. (e) Photoemission intensity along the vertical dashed line in panel a for original data, the reconstruction from the full-scale image and the reconstruction from the zoomed-in image. (f) Corresponding intensity curves for the horizontal dashed line in panel a.}\label{fig:4}
\end{figure}

To explore this in detail, Figs.~\ref{fig:4}(a) and~(b) show the same data as the first column of Fig.~\ref{fig:3}, presenting the lower-quality reconstruction of the sharpest features more clearly. We test ARPESNet on a zoomed-in selection of the image centre. Figs.~\ref{fig:4}(c) and (d) show the selected central region and its reconstruction. We note that the reconstruction is significantly improved, as judged by visual inspection. To show this improvement quantitatively, Figs.~\ref{fig:4}(e) and~(f) give profiles of the photoemission intensity in the original and reconstructed images along the vertical and horizontal dashed lines in panel~(a). While the reconstruction of the full image is rather poor, the reconstruction of the zoomed-in selection agrees very well with the original. To understand the apparent shortcomings of the full image's reconstruction, one needs to keep in mind that the training and test data is arbitrarily scaled. As already mentioned, an overly complex input image is unlikely to be reconstructed well. Indeed, in the extreme case of features with a size similar to the pixel separation in the image, these would not even be well-represented in the input image. Such overly complex images could result, e.g., by extending the~$k$-range of the input data over many Brillouin zones, such that many closely separated features are present in the image. Sharp structures in the spectra lead to a similar situation, the very narrow and intense bands on Bi(111) being such a case.

\section{Results}

Our overall objective is to craft a compressed representation of ARPES data in latent space that extracts the most important spectral features and enables efficient further on-the-fly analysis. To assess the quality of latent-space representations, we test how well these can be clustered by $k$-means. We first explore clustering vs. a known ground truth and compare this performance to that for raw images, as well as to images compressed using DCT22. We then study clustering in the situation of actual experimental data from a graphene device.

\begin{figure}[htbp]
\centering
\includegraphics[width=0.8\columnwidth]{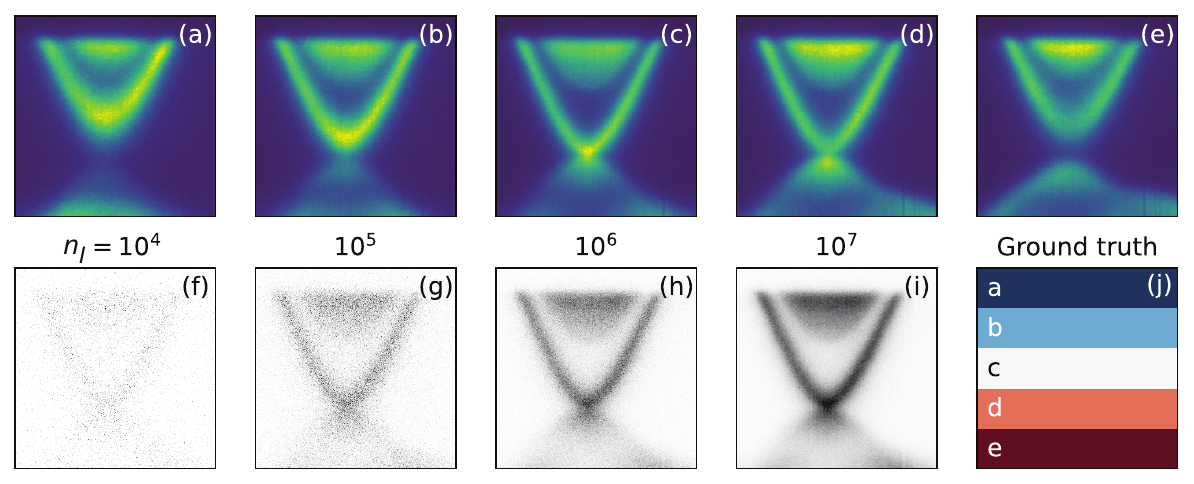}
\caption{Model for testing the suitability of the compressed data for clustering. (a)-(e) Five slightly different ARPES spectra from Bi$_2$Se$_3$ \cite{Bianchi:2010ab}. The spectra are taken from the reference dataset and have a high $S/N$. (f)-(i)~Noisy spectra derived from the original data in panel~c assuming a different total number of counts~$n_I$ per image; these are displayed in greyscale, which makes it easier to see structures for low~$n_I$. (j)~Ground truth for clustering; each coloured stripe has~500 pixels that correspond to~500 different noisy spectra, all derived from one of the five reference spectra. A successful $k$-means clustering must produced a permutation of these coloured stripes.}
\label{fig:5}
\end{figure}

For the performance test, it is desirable to define a ground truth while at the same time retaining relevance to the actual situation encountered in experiments. We do so by the following procedure: We first choose an arbitrary number of different high-quality spectra (in this case, five). Figs.~\ref{fig:5}(a)-(e) show this set. All spectra show the surface state of Bi$_2$Se$_3$ \cite{Bianchi:2010ab}. They are quite similar with the only difference that not all cuts in $\mathbf{k}$ perfectly hit the Dirac point in the surface state dispersion (only panel (c) does).
From each image in such a dataset, we generate $l$ different noisy versions ($l = 500$) as described above, for a pre-defined total of $n_I$ counts in the image. Examples of the same spectrum with different values of $n_I$  are given in Fig. \ref{fig:5}(f)-(i). We see that $n_I=10^7$ counts gives a very high $S/N$ whereas at $n_I=10^4$ counts it would be difficult to distinguish similar images, e.g., images derived from panels (c) and (d), by eye. Evidently, the ability to distinguish between the spectra at a high noise level strongly depends on how different the five spectra are from each other. Here we have deliberately chosen an example with relatively similar spectra.

For a given electron count number $n_I$, we then use $k$-means to cluster all the $5 \times 500 = 2,500$ spectra into five categories and we subsequently evaluate the quality of this clustering by comparing the category resulting from $k$-means to the original category the spectrum is derived from, i.e., to the ground truth. Ideally, all noisy spectra are correctly categorised, resulting in 500 spectra in each $k$-means cluster. As the total electron count number / acquisition time decreases, we expect the fraction of incorrectly assigned images to increase.

In order to visualise the quality of the clustering, we can think of the ground truth as the two-dimensional ``sample'' in Fig. \ref{fig:5}(j) which consists of five stripes, each containing 500 points, representing one of the five spectra shown in Fig. \ref{fig:5}(a)-(e). Ideally, the $k$-means clustering of the noisy data should be able to reproduce this stripe structure (actually, just a permutation of the stripe structure because the cluster labels produced by $k$-means are arbitrary). Incorrectly clustered spectra can easily be recognised by several stripes having the same colour or by dots  of the wrong colour / mixing of the clusters. This simple visualisation actually corresponds to a realistic situation in a nanoARPES experiment in which different electronic structures and thus different ARPES spectra might be present in well-separated domains on a sample surface. Striped domains of the kind shown here could be expected for a Bi$_2$Se$_3$ sample with a curved surface or domains of different crystalline orientation \cite{Agustsson:2024aa}.

\begin{figure}[htbp]
\centering
\includegraphics[width=1.0\columnwidth]{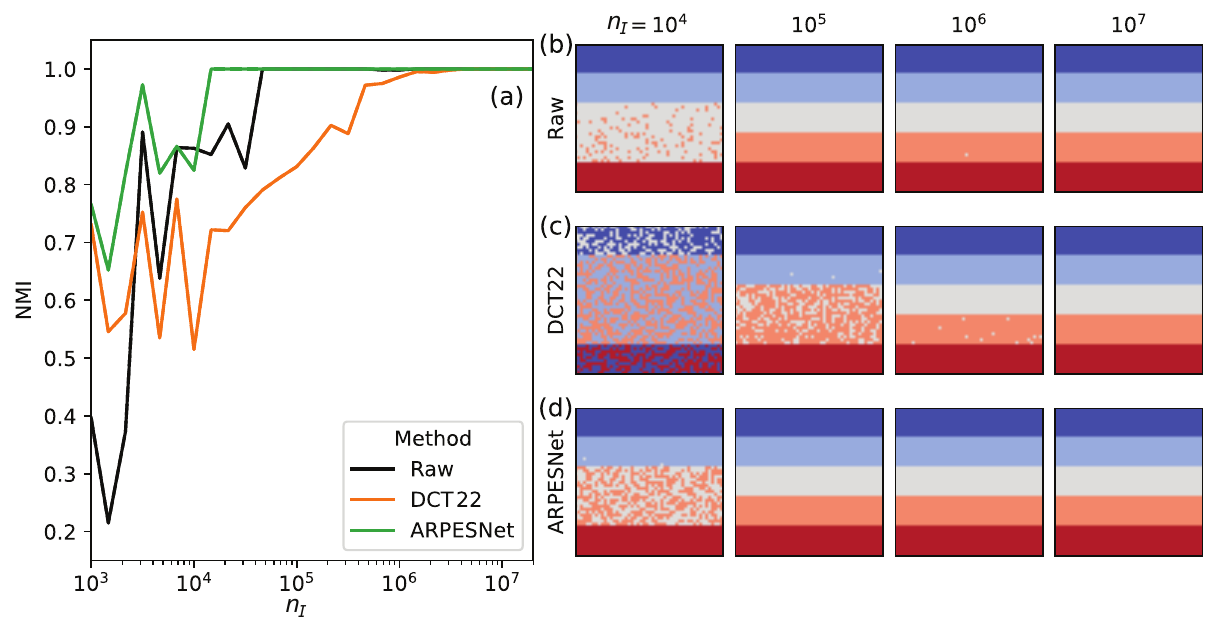}\\ 
\caption{Clustering performance of ARPESNet compared to DCT22 and to no compression. 
(a)~Normalized mutual information (NMI) of as a function of total counts per image~$n_I$. 
(b)-(d)~Visualization of the clustering result for the three data representations for four values of~$n_I$ each. Perfect clustering corresponds to the image of Fig.~\ref{fig:5}(j) and is achieved for all data representation at high count rates. For visual aid, the colours assigned to the clusters are the permutation of the cluster labels which most resembles the ground truth.}\label{fig:6}
\end{figure}

Fig. \ref{fig:6}(a) shows the clustering performance for the test data set in Fig.~\ref{fig:5} as a function of total counts $n_I$ per spectrum.  The clustering was carried out with $k=5$, as in the ground truth, picking the best of 100 random initialisations. As a performance measurement, we use the normalised mutual information (NMI) \cite{Knops:2006aa} for different compression techniques. We show results for clustering uncompressed spectra (raw), spectra compressed with DCT22 and finally spectra compressed by ARPESNet.
Figs.~\ref{fig:6}(b)-(d) illustrate the clustering result at different noise levels for the three representations using the visualisation of Fig.~\ref{fig:5}(j).

Clustering performance is perfect for~$n_I \geq 10^7$, not only for the raw data, but also for the two compression techniques. As~$n_I$ is lowered, clustering performance deteriorates for all three techniques. DCT22 is the first to show deterioration with NMI$<1.0$ at~$n_I < 4 \times 10^6$, followed by uncompressed clustering at $n_I<6\times10^4$. ARPESNet shows perfect clustering until $n_I<2\times10^4$, while continuing to outperform the two other data representations at lower counts, even though no longer performing reliably. It is particularly remarkable that ARPESNet outperforms clustering on the raw data, especially for the lowest $n_I$ tested. 

\begin{figure}[htbp]
\includegraphics[width=1.0\columnwidth]{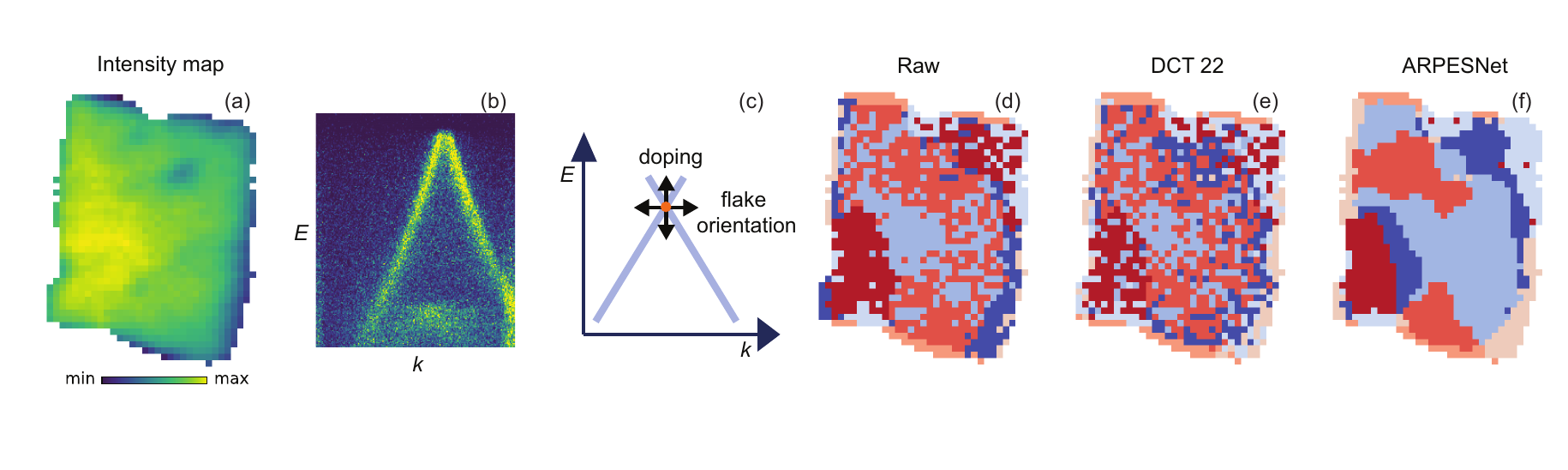} 
\vspace{-2mm}
\caption{Clustering performance on a real microARPES raster map with data from Ref. \cite{Curcio:2020aa}. (a) Sum intensity of the photoemission signal as a function of measured position. (b) A single spectrum from one of the pixels in panel (a). (c) Dirac cone movements in $k$ are caused by rotational orientation changes of the flake. Movements in energy are caused by different doping levels. (d)-(e) Map of clustering labels obtained with $k$=8 on raw data,  DCT22 compressed data and  data compressed using ARPESNet, respectively.}\label{fig:7}
\end{figure}

We now test $k$-means clustering with the two compression techniques on an actual microARPES data set, collected on an electronic device made of graphene.  In this experiment, the exciting UV spot for ARPES is scanned across the surface of a graphene flake and an ARPES image is measured at each position. The details of the experiment and the data analysis are described in Ref.~\cite{Curcio:2020aa}. Here we explain the basic data features and explore the clustering quality using compressed and raw data. Fig.~\ref{fig:7}(a) shows a real-space photoemission image of the graphene flake, where each pixel corresponds to an ARPES image as those in Fig.~\ref{fig:7}(b); the intensity in Fig.~\ref{fig:7}(a) is integrated over the entire spectrum (see Fig.~2(e) in Ref.~\cite{Curcio:2020aa}). This total photoemission intensity is of limited use as it lacks a direct physical interpretation. Nevertheless, it is easy to generate while the experiment runs and nanoARPES experiments are often guided by such intensity images. The individual spectra for each pixel contain much information about the system. Fig.~\ref{fig:7}(b) shows such a spectrum with the typical Dirac cone of graphene. In different images across the sample, this cone-like structure may be shifted in energy or~in $k$, as indicated in Fig.~\ref{fig:7}(c).  These shifts have physical interpretations as change of doping and change of graphene flake orientation, respectively (see Fig. 3 in Ref. \cite{Curcio:2020aa}).

Figs.~\ref{fig:7}(d)-(f) show the result of clustering the individual spectra in the graphene flake. The data are represented via the same pixels as in Fig.~\ref{fig:7}(a) but now pixel colours stand for cluster numbers resulting from $k$-means. The cluster positions are similar for all three representations, yet ARPESNet clearly outperforms the other two, yielding a map with significantly less noise. Comparing the cluster landscape to a conventional analysis in Fig.~3 of Ref.~\cite{Curcio:2020aa}, it becomes clear that the $k$-shift of the cone in the electronic structure is the dominant factor in clustering and clusters found by $k$-means denote areas of slightly different graphene orientation. It makes little physical sense that this orientation would locally fluctuate on a small scale, as different orientations are often caused by domain boundaries formed by folding the graphene sheet. The low-noise representation delivered by ARPESNet is thus most meaningful from a materials point of view.

\begin{figure}[htbp]
\centering
\includegraphics[width=0.5\columnwidth]{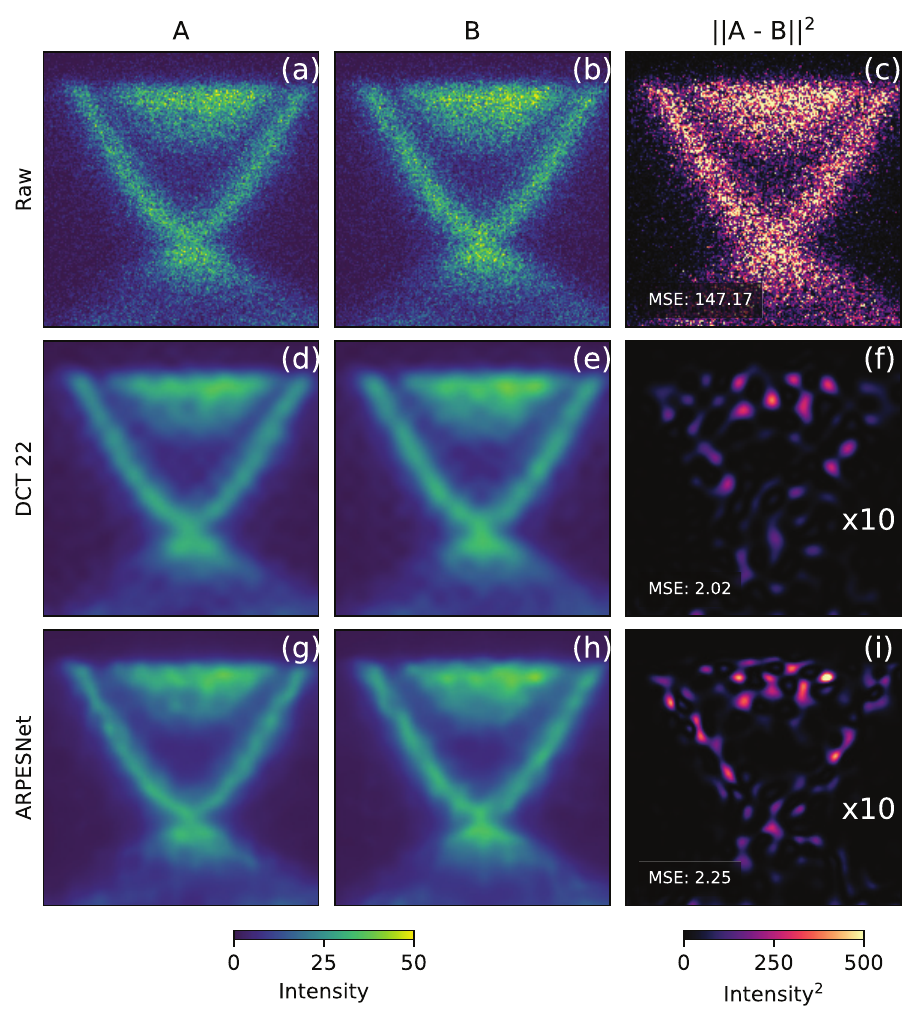} 
\caption{(a), (b) Two versions of the same input image, augmented with different random noise  ($n_I=10^5$). (c) Pixel-wise squared difference between these images. (d)-(f) and (g)-(i) Corresponding images and difference when using the output images of DCT22 and ARPESNet respectively. The values in (f) and (i) have been multiplied by a factor 10 and are displayed with the same colour scale as panel c.}\label{fig:8}
\end{figure}

Why does ARPESNet perform so relatively well on clustering noisy data? To address this question, we first show why one should expect poor clustering performance on the raw data. To this end we compare two \emph{different} noisy versions ($n_I=10^5$) of the same instance in the training set in Figs. \ref{fig:8}(a) and (b), as well as their pixel-wise squared difference in Fig. \ref{fig:8}(c). Although images~A and~B looking essentially the same, the squared difference is high (MSE = 157.41) and shows the same structure as the raw images. This result is due to the pixelated character of noise at low $S/N$: ARPES images show the photoemission intensity generated by electrons hitting the detector. In the limit of a very low total number of counts, the images are dominated by single count events. The density of these counts is larger in some regions, yet, due to the random location of those regions, a pixel-wise comparison of two noisy images derived from the same original image shows a significant difference. With this fact in mind, it is plausible that the clustering performance on raw data is low.

Figs.~\ref{fig:8}(d)-(f) and~(g)-(i) show the corresponding comparison after processing images~A and~B using DCT22 or ARPESNet, respectively. The agreement between the processed images is much better, with MSE(DCT22)=2.02 and MSE(ARPESNet)=2.25. It is easy to see why: Both compression techniques lead to smoothing and the ensuing mitigation of the pixelated character causes MSE to drop sharply. Interestingly, DCT22 attains a slightly lower MSE than ARPESNet. Still, the images reconstructed by DCT22 in Figs. \ref{fig:8}(d) and (e) exhibit a weak but distinct grid-like structure that is absent in the original data and in the images reconstructed by ARPESNet.

Considering the clustering performance on the compressed images, we stress that $k$-means is performed on latent-space representations, not on the reconstructed images in Fig.~\ref{fig:8}. Therefore, no conclusions about the clustering characteristics can be strictly inferred from Fig.~\ref{fig:8}. Still, it is highly plausible that using the compressed representations outperforms the raw data simply because the pixelated noise cannot be represented in the compressed space. Understanding why clustering using ARPESNet outperforms clustering by DCT22 is more difficult. We speculate that this could be related to how the data is represented in the latent space. DCT is designed to capture smoothly variegated images via a hierarchy of predefined cosine functions at progressively finer resolution; on the other hand, ARPESNet provides a more powerful and versatile neural representation attuned to the data. We thus speculate that, even if DCT22 attains lower reconstruction error, ARPESNet learns to represent salient features of images that are more consequential for clustering purposes.

While these considerations may explain the excellent suitability of data processed by ARPESNet for clustering, especially compared to raw data, the denoising capability of ARPESNet is still remarkable, especially as it has not been explicitly trained for denoising. Nevertheless, the training with many different noisy images combined with the limited size of the latent space results in a high denoising performance.

\section{Conclusion}

We have constructed an autoenconder network, ARPESNet, with an architecture and hyperparameters suitable for treating ARPES images. ARPESNet generates a highly compressed version of ARPES data in latent space, by a factor of~128, and yields a high-quality reconstruction by decoding. To train this network, we proposed a way to extract large quantities of training data from three-dimensional ARPES datasets by the placement of random cuts in~$(k_x, k_y)$ space. Such a data extraction strategy could be used for all kinds of three dimensional datasets, including medical computed tomography and magnetic resonance imaging scans. Despite the high compression, $k$-means clustering of ARPES images compressed by ARPESNet outperforms that on the raw data in quality, both on synthetic data with a known ground truth and on real data from a graphene flake device. We ascribe this high-quality clustering performance attained by ARPESNet to the denoising effect obtained when training the network with noisy data. It is feasible to integrate data compression by ARPESNet in the data collection process, such that one ARPES image is compressed while the next is measured, especially since the model does not require re-training when investigating different materials. In this way, the strongly compressed data representation can be made available for data analysis on the fly.

\section{Acknowledgement}

This work was supported by a research grant (40558) from VILLUM FONDEN.

\section{Code and data availability}

The code for ARPESnet is available on github at https://github.com/ARPES-on-the-fly/arpesnet. The data for training and testing ARPESNet is available on Zenodo (DOI: 10.5281/zenodo.1264878). The entire ARPES data base used for deriving the training and testing datasets is also available on Zenodo (DOI: 10.5281/zenodo.12665275). Please note that the Zenodo DOIs will first become functional after the final publication of this paper. In the meantime, the data is available from the authors upon request.

\vspace{5mm}


\end{document}